# Structural and Optical Properties of Crystal Ion Sliced BaTiO$_3$ Thin Films

H. Esfandiar[1,2,*], F. Abtahi[2], T. G. Vrckovnik[1,2,3], G. Q. Ngo[1,2], R. Heller[4], U. Kentsch[4], F. Ganss[4], S. Facsko[4], U. Lucchesi[4], S. Winnerl[4], F. Eilenberger[1,2,3], D. Arslan[1], S. W. Schmitt[1,2,+]

[1] *Fraunhofer Institute for Applied Optics and Precision Engineering IOF, Albert-Einstein-Str. 7, 07745 Jena, Germany*

[2] *Institute of Applied Physics, Abbe Center of Photonics, Friedrich Schiller University Jena, Albert-Einstein-Str. 15, 07745 Jena, Germany*

[3] *Max Planck School of Photonics, Albert-Einstein-Str. 15, 07745 Jena, Germany*

[4] *Institute of Ion Beam Physics and Materials Research, Helmholtz-Zentrum Dresden-Rossendorf, 01328 Dresden, Germany*

[*]hossein.esfandiar@iof.fraunhofer.de

[+]sebastian.wolfgang.schmitt@iof.fraunhofer.de

**Barium titanate (BaTiO$_3$) is a compelling material for integrated photonics due to its strong electro-optic and second-order nonlinear properties. Crystal Ion Slicing (CIS) presents a scalable and CMOS-compatible route for fabricating thin BaTiO$_3$ films; however, ion implantation during CIS introduces lattice damage that can degrade structural and optical performance. In this study, we demonstrate that post-slicing thermal annealing effectively restores the structural integrity and optical quality of CIS-processed BaTiO$_3$ flakes. Raman spectroscopy confirms the recovery of crystallinity, while second-harmonic generation (SHG) microscopy reveals systematic reorientation of ferroelectric domains and restoration of the associated second-order nonlinear susceptibility tensor, $\chi^{(2)}$. Notably, SHG signals persist even in regions with weak Raman signatures, indicating that long-range ferroelectric order can survive despite partial lattice disruption. Optical measurements show that the linear dispersion of annealed CIS flakes closely matches that of bulk BaTiO$_3$, validating their suitability for photonic integration. Together, these results qualify CIS – combined with thermal annealing – as a viable and scalable manufacturing strategy for high-quality BaTiO$_3$-on-insulator (BTOI) platforms, enabling advanced integrated photonic devices for modulation, frequency conversion, and quantum optics.**

## Introduction

BaTiO$_3$ is a ferroelectric perovskite oxide that has gained significant attention as an electro-optic and nonlinear material in quantum optics due to its exceptional optical properties[1,2]. Its high electro-optic coefficient facilitates efficient light modulation, while its substantial nonlinear susceptibility makes it suitable for applications such as frequency conversion and optical switching[3,4]. As the demand for integrated optical systems grows across various fields, including telecommunications and quantum computing, BaTiO$_3$ emerges as an ideal candidate for a wide range of applications, including integrated resonators, modulators, switches, and frequency converters[3–10].

To date, the fabrication of crystalline BaTiO$_3$ for these applications mostly relies on molecular beam epitaxy (MBE)[11] or MBE buffer deposition in combination with high temperature RF sputtering for epitaxial film growth on Si substrates[4]. While effective in producing high-quality thin films, those methods are often constrained by complexity, lengthy processing times, limited film thicknesses of a few 100 nm and relatively high deposition costs. In addition, for most integrated optical applications subsequent bonding of the thin films on SiO$_2$ / Si substrate and handler wafer removal is needed. These limitations hinder the scalability and accessibility of BaTiO$_3$-based devices, highlighting the need for alternative fabrication methods that enhance production efficiency and reduce costs.

Crystal ion slicing (CIS) has emerged as an alternative technique for fabricating BaTiO$_3$ thin films[12,13]. CIS simplifies processing and enables high-quality bonding of single-crystal layers onto a wide range of substrates, facilitating their integration into optical systems[14,15]. However, the influence of ion-induced damage on the structural, linear, and nonlinear optical properties of demonstrably sensitive ferroelectric BaTiO$_3$ films remains insufficiently understood. Such damage degrades crystallinity and accordingly alters refractive indices, increase optical losses, and impairs nonlinear optical performance. These effects would critically limit the suitability of CIS-fabricated BaTiO$_3$ films for applications in integrated photonics, nonlinear optics, and electro-optic devices, where high material quality and low optical loss are essential.

In this study, we demonstrate that CIS and post slicing annealing can produce BaTiO$_3$ films that exhibit linear and nonlinear optical properties comparable to bulk BaTiO$_3$. This confirms that CIS enables the fabrication of high-quality BaTiO$_3$ films suitable for advanced photonic integration.

## Results and discussion

### Ion implantation and investigation of the as-implanted BaTiO$_3$ crystal

To prepare BaTiO$_3$ thin films with thicknesses between 1.1 and 1.3 µm by ion slicing, that recently were identified as suitable for integrated quantum optics waveguide fabrication[10], two (001)-oriented (out of plane c-domain polarized), polished BaTiO$_3$ single crystals were implanted with 480 keV He$^+$ ions. Ion fluences of $5 \times 10^{16}$ ions/cm² and $2 \times 10^{17}$ ions/cm² were used, referred to as 'low fluence' and 'high fluence' throughout the manuscript, respectively. SRIM (Stopping and Range of Ions in Matter, Fig. 1a) simulations of He$^+$ implantation reveal a maximum of the projected ion range R$_p$ at approximately 1.1 µm depth with a straggling of about 200 nm (FWHM) within the BaTiO$_3$ target in accordance with the targeted film thickness.

Non-polarized Raman spectroscopy in backscattering configuration was used to investigate the structure of pristine, as-implanted low-fluence and high-fluence BaTiO$_3$ crystals (Fig. 1b). The most prominent peaks in the Raman spectrum of the pristine crystal are marked by dashed vertical lines and correspond to characteristic lattice vibrations of bulk tetragonal BaTiO$_3$[16,17]. Typically, those peaks can be assigned to the Raman active modes A$_1$(TO) (~175 cm$^{-1}$, ~270 cm$^{-1}$, ~520 cm$^{-1}$), B$_1$ or E(TO+LO) (~305 cm$^{-1}$), A$_1$(LO) (~470 cm$^{-1}$, 720 cm$^{-1}$) that can partially be convoluted with other modes in the presented case of an experimental setting that is not polarization-selective[18,19]. The appearance of the B$_1$ mode at ~305 cm$^{-1}$ directly indicates tetragonality, as it is forbidden in the centrosymmetric cubic phase and becomes Raman-active only when symmetry is broken in the tetragonal phase[20]. Additionally, mode splitting of previously degenerate phonons (e.g., T modes) into A$_1$, E, and B$_1$ components reflects the lower symmetry of the tetragonal phase[16,17]. Together, these spectral features unambiguously signal a tetragonal distortion of the BaTiO$_3$ lattice of the as-received BaTiO$_3$ crystal at room temperature.

Following He$^+$ ion implantation, a few significant changes are observed in the Raman spectrum (see spectra 'low' and 'high' in Fig. 1b). In particular, the relative decrease of the B$_1$ (~305 cm$^{-1}$) mode compared to the A$_1$(TO) (~270 cm$^{-1}$) mode with increasing ion fluence points towards the expected structural degradation and a loss of tetragonality (asymmetry of the crystal along the c-direction) for the irradiation-damaged crystal. The B$_1$ mode is symmetry-allowed only in the tetragonal phase and its suppression reflects the restoration of local centrosymmetry and the collapse of the longer-range order. The overall increase of A$_1$(TO) modes, accompanied by the disappearance of A$_1$(LO) modes already at low ion fluence, can be explained by a reorientation of domains from out-of-plane (c-domains) of the pristine crystal to in-plane (a-domains) in the implanted surface layer. This domain reconfiguration alters the Raman selection rules, suppressing LO modes (which are strongly excited in z-polarized, backscattering geometry) and enhancing TO modes (which dominate when the polarization lies in-plane)[21,22]. This polarization flip is likely driven by strain fields and defect gradients introduced by ion implantation. In suspended BaTiO$_3$ films without mechanical constraints, these effects can energetically favor in-plane polarization as a means to relieve stress – a phenomenon known in the literature as 'strain doping'[23]. Different results were found for BaTiO$_3$ films by pulsed laser deposition, where a mixed domain configuration changed to c-domain[24]. At the high implantation fluence, even the $A_1$(TO) modes begin to decrease, indicating more severe disruption of the lattice. Point defects, interstitials, and He-related defect complexes introduced by the implantation act as phonon scattering centers, reducing the coherence and lifetime of optical phonons. This leads to mode broadening, intensity loss, and eventually, a transition toward a quasi-amorphous or heavily disordered state, where longer-range polar order and phonon selection rules break down.

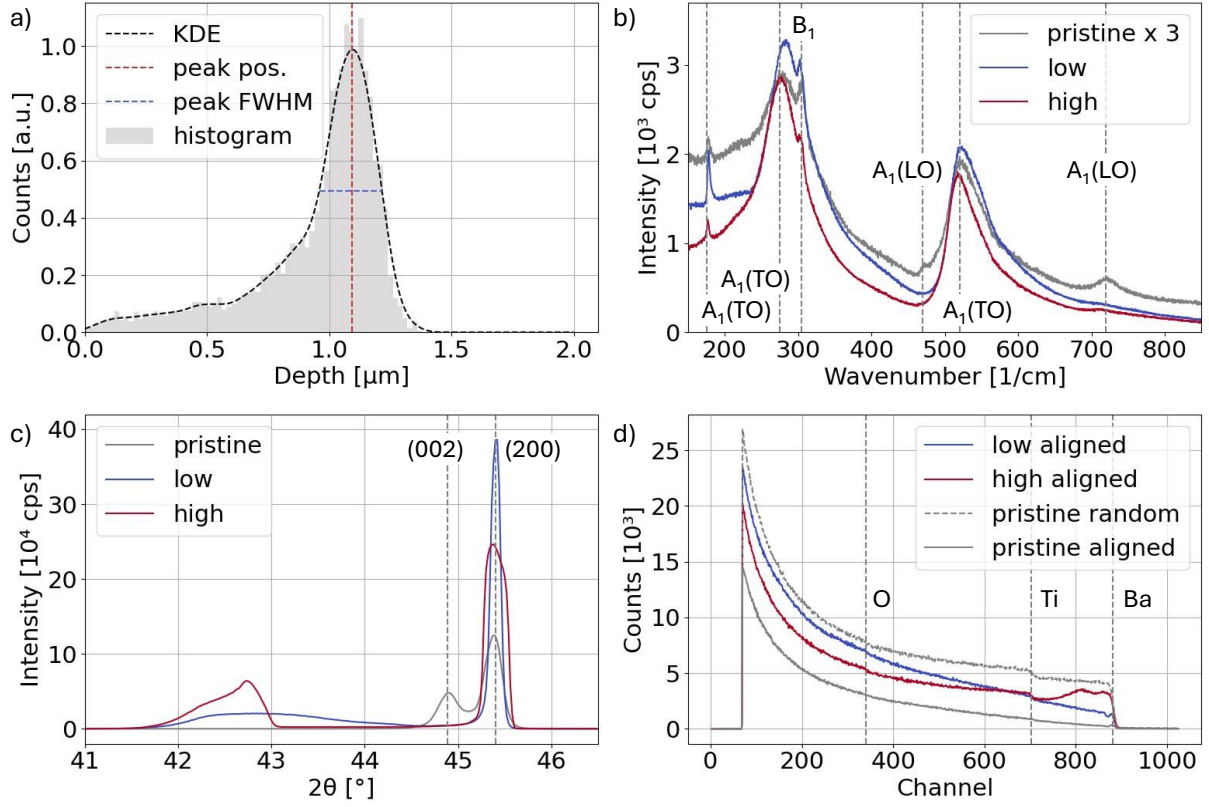

**Figure 1. Ion implantation and structural analysis of pristine and as-implanted BaTiO$_3$ crystals. a)** Projected ion range histogram with kernel density estimation (KDE / dashed black line) determined by a SRIM simulation of He$^+$ ions (450 keV) implanted in BaTiO$_3$ vs. depth. Red and blue dashed lines show position and FWHM of the histogram peak, respectively, indicating damage maximum and straggle. **b)** Raman spectra (the dashed vertical lines show BaTiO$_3$ characteristic Raman peaks), **c)** XRD 2θ/ω scan (dashed vertical lines indicate expected (002) and (200) reflections, respectively) and **d)** RBS spectrum (dashed vertical lines indicate the onset of Ba, Ti, O backscattering respectively) of a pristine, low (5·10$^{16}$ ions/cm$^2$) and high fluence (2·10$^{17}$ ions/cm$^2$) He$^+$ ion implanted BaTiO$_3$ bulk crystal surface with 001-orientation (c-axis out of plane).

Fig. 1c presents the high-resolution X-ray diffraction (XRD) pattern of the BaTiO$_3$ crystals before (pristine) and after implantation with high and low He$^+$ fluence. The sharp, well-defined peaks for the pristine crystal align with the expected reflections of the tetragonal perovskite structure (space group P4mm). Hallmarks of this phase are particularly prominent in the 2θ range of 44° - 46°, where the (002) and (200) reflections are clearly distinguished. This splitting results from the anisotropy in lattice parameters (c ≠ a) inherent to the tetragonal symmetry, in contrast to the single peak expected for the cubic phase[25]. The (002) reflection – associated with the c-axis – appears at 44.9°, while the (200) reflection – corresponding to the a-axis – occurs at 45.4°. The reflection peak separation provides a direct measure of tetragonality, which reflects the degree of ferroelectric distortion in BaTiO$_3$. The observed splitting thus confirms that the as-received crystal is both structurally well-ordered and retains strong ferroelectric character.

Following He$^+$ ion implantation, notable modifications appear in the diffraction pattern. While the (200) reflection remains visible across all implanted samples and retains its original position, the width of this peak increases slightly for the high implantation fluence. This broadening is indicative of enhanced lattice disorder and microscopic strain, likely due to implantation-induced point defects and residual stress. In contrast, the (002) reflection not only broadens but also shifts notably toward lower angles, suggesting a localized expansion along the c-axis in the implanted region that can be associated with interstitial or local accumulation of $He^+$. This asymmetric behavior between the (002) and (200) reflection indicates the development of anisotropic strain fields within the damaged lattice.

Moreover, the pronounced relative increase in intensity of the (200) reflection following implantation of the lower ion fluence provides additional evidence for a preferential enhancement of a-domain orientation in the near-surface region. This observation is consistent with the Raman spectroscopy results, which indicate a suppression of vibrational modes associated with the out-of-plane polarized c-domains. In parallel, the pronounced broadening and low-angle shift of the (002) reflection strongly suggests that the c-domain lattice component has undergone substantial structural disruption.

Fig. 1d presents the Rutherford Backscattering Spectrometry (RBS) channeling spectra obtained from $BaTiO_3$ crystals implanted with low and high helium ion fluences. The random and aligned (channeling) spectra of the pristine $BaTiO_3$ crystal were recorded to evaluate the extent and depth distribution of ion-induced lattice damage.

As evident from the spectra, a broad damage-related enhancement in the backscattered yield emerges near the surface region, becoming significantly more pronounced at higher implantation fluence. For the high-fluence sample, a distinct damage peak is observed, reflecting a higher density of displaced atoms and extended defects. This increase in RBS yield is a direct indicator of increased lattice disorder, as the loss of crystallographic order allows more ions to be backscattered rather than channeled through the crystal lattice. The observed fluence dependence of the damage signal is consistent with the collision cascade model, wherein each incident $He^+$ ion initiates a series of atomic displacements that propagate through the lattice[26]. At higher fluences, the overlap of cascades leads to enhanced defect accumulation, vacancy clustering, and potentially the formation of extended defects such as dislocation loops and helium-vacancy complexes. These extended defects significantly increase the dechanneling probability, thereby amplifying the RBS yield in the aligned configuration. Further, the RBS results are in good agreement with previous studies on, e.g., Fe implantation in $BaTiO_3$[27].

**$BaTiO_3$ flake exfoliation**

Following $He^+$ ion implantation, the surface morphology of the $BaTiO_3$ crystal remains largely unchanged, with no visible damage or blister formation. This is demonstrated in Fig. 2a, which shows a bright field microscopy image of the surface of the high-fluence implanted $BaTiO_3$ crystal. Such an intact surface appearance is expected, as most of the implantation-induced damage occurs beneath the surface, localized around the projected ion range $R_p$. The overall structural integrity of the surface post-implantation confirms that the implantation fluence and energy were well-controlled, avoiding premature blistering or delamination.

Upon post-implantation annealing significant subsurface microstructural changes occur. The implanted helium atoms, initially present as interstitials or small clusters, become mobile and coalesce into nanoscale bubbles[28]. As these bubbles grow, they generate localized pressure, creating mechanical stress at the interface between the damaged layer and the underlying bulk crystal.

With continued annealing or mechanical assistance (e.g., bonding to a handle substrate), this stress can induce microcrack formation and delamination along the helium-rich plane. The resulting cracks propagate laterally, enabling clean, controlled exfoliation of a thin, single-crystalline $BaTiO_3$ layer defined by the implantation depth and retaining its original crystallographic orientation.

Figure 2b illustrates the process, showing the high-fluence implanted $BaTiO_3$ surface after transfer onto a $Si/SiO_2$ (300 nm) substrate and subsequent heat treatment on a hotplate at 270 °C for 1 hour to induce blistering and exfoliation. The detachment of flakes leads to distinct topographical steps, visible at the center of the image. For AFM analysis, a specific region – highlighted by the yellow rectangle in Fig. 2b – was selected to directly measure across the flake-substrate boundary, enabling precise determination of the exfoliation-induced step height.

The corresponding AFM topography profile, shown in Figs. 2c and d, reveals a well-defined step with a thickness of approximately 1.2 μm. This value aligns well with the mean projected ion range $R_p$ and straggling predicted by SRIM simulations for 480 keV helium ion implantation, as shown in Fig. 1a. The close agreement between the AFM measurements and the simulated ion penetration depth strongly supports the conclusion that cleavage occurred precisely along the ion-induced damage layer.

The root mean square (RMS) roughness values derived from the AFM image in Fig. 2c – taken from the $BaTiO_3$ surface after blistering (right) and after exfoliation (left) – range between 10–20 nm. This indicates that the CIS process, under the applied conditions, yields $BaTiO_3$ flakes with reasonably smooth surfaces on both sides.

The process produced $BaTiO_3$ flakes of varying lateral sizes on the substrate, as exemplarily shown in the optical micrograph in Fig. 2e. Notably, significant delamination was observed only in the high-fluence samples, suggesting that the lower fluence was insufficient to induce the desired exfoliation. Consequently, all further investigations focus exclusively on flakes obtained from the high-fluence samples.

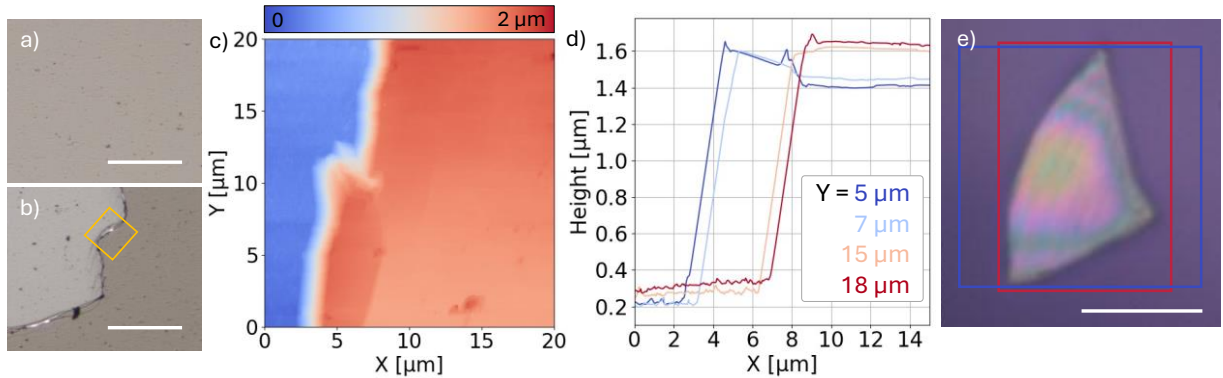

**Figure 2. $BaTiO_3$ blistering and flake exfoliation. a)** Bright field microscopy image of **a)** as implanted $BaTiO_3$ surface with high fluence ($2 \cdot 10^{17}$ ions/cm$^2$), scale bar is 40 μm and **b)** same surface after high temperature blistering and removal of parts of the surface thin film, scale bar is 40 μm. The yellow rectangle shows the area of the AFM topography scan in c). **c)** AFM topography of surface area indicated in b). **d)** Topography profiles of the AFM map in c) in various heights. **e)** Bright field optical microscopy image of a $BaTiO_3$ flake exfoliated on a 300 nm $SiO_2$ thin film on Si. The red and blue frames show the area of the Raman and SHG microscopy scan in Figs. 3 and 4, respectively. The scale bar is 10 μm.

**$BaTiO_3$ flake defect curing, nonlinear and linear optical investigation**

To promote structural healing of ion-induced damage in the exfoliated $BaTiO_3$ flakes annealing was performed in a conventional furnace at 700 °C for 21 hours in ambient air. To monitor the evolution of the $BaTiO_3$ crystal structure before and after annealing, spatially resolved Raman spectroscopy was employed. K-means clustered Raman maps of the $BaTiO_3$ flake shown in Fig. 2e (red frame) prior to and following annealing are presented in Fig. 3a (left and right, respectively)[29]. The corresponding Raman centroid spectra (Fig. 3b) reveal notable spectral changes indicating substantial recrystallization and potential domain reconfiguration induced by thermal treatment.

Specifically, the pronounced enhancement of the $A_1(TO)$ mode intensity after annealing signifies the recovery of the disrupted perovskite lattice and a partial restoration of tetragonal symmetry. Additionally, the relative increase of the $B_1$ mode intensity compared to $A_1(TO)$ further suggests improved structural ordering and strengthening of tetragonality.

The intensification of one $A_1(LO)$ mode at 720 cm$^{-1}$ post-annealing may reflect a reorientation of some domains from in-plane (a-domains) towards out-of-plane (c-domain) configurations, reminiscent of pristine bulk BaTiO$_3$. However, the overall dominance of $A_1(TO)$ modes in the spectra of the crystal flakes implies an overall in-plane alignment of the ferroelectric domains (a-domain).

Notably, lattice recovery appears to initiate at the edges of the flake and progress toward the center. This behavior may be attributed to the reduced mechanical confinement at the edges and more efficient helium outgassing and defect mobility due to the higher surface-to-volume ratio in these regions.

A similar behavior for Raman mapping before and after annealing is found for a second BaTiO$_3$ flake as can be seen in Supplementary Fig. S1.

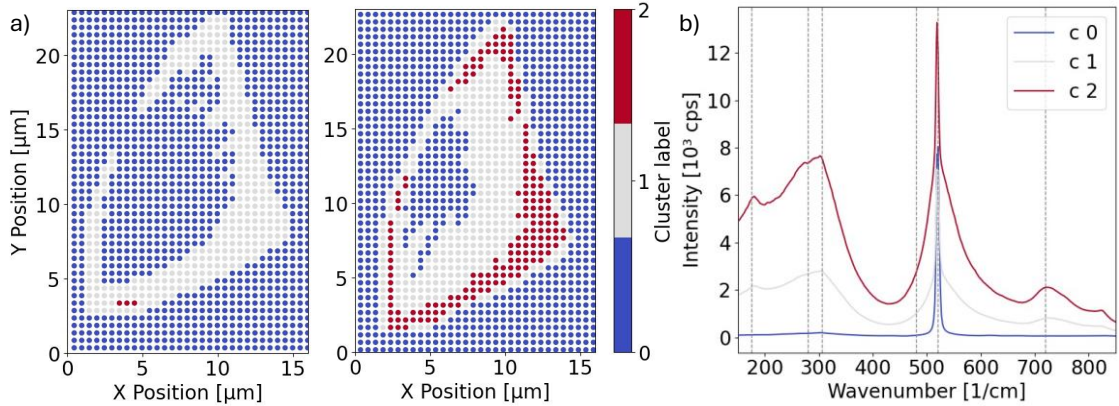

**Figure 3. Raman mapping of BaTiO$_3$ flake before and after annealing. a)** Clustered Raman maps before (left) and after (right) annealing of the BaTiO$_3$ flake shown in Fig. 2e. **b)** Cluster centroid spectra of the Raman maps shown in a). Dashed vertical lines show BaTiO$_3$ characteristic Raman peaks.

Domain reorientation before and after annealing of the BaTiO$_3$ flakes can be better understood by polarization-resolved second-harmonic generation (SHG) mapping. It exploits the nonlinear optical response of non-centrosymmetric materials like BaTiO$_3$ to probe ferroelectric domain structures. When illuminated with polarized light, the material generates light at twice the frequency (second harmonic), with an intensity that depends on the local orientation of the nonlinear susceptibility tensor. Since each ferroelectric domain has a distinct polarization direction, it produces a unique SHG signal pattern as the input polarization is varied. By scanning the polarization and detecting SHG intensity changes, one can map domain orientations with high spatial resolution. This non-destructive technique provides insight into domain alignment, domain boundaries, and symmetry properties.

Polarization-resolved SHG microscopy mapping was carried out in a backscattering configuration both before and after thermal annealing, focusing on the blue-framed region highlighted in Fig. 2e. For data evaluation, it was assumed that the extraordinary axis of the BaTiO$_3$ flake crystal lies predominantly within the plane of the sample surface (see Methods for a detailed description of the experimental setup). The resulting SHG maps are presented in Fig. 4a, where the relative SHG intensity is represented by brightness, and the in-plane ferroelectric domain orientation (ranging from 0° to 180°) is encoded by color. The orientation values were extracted by fitting the polarization-resolved SHG signals acquired at each pixel (see Supplementary Information for the data fitting procedure). To illustrate the fitting quality and variation in domain orientation, Fig. 4b shows three representative polarization-dependent SHG intensity curves, each corresponding to a distinct location marked in Fig. 4a. We see strong reorientation of the in-plane domain orientation before and after annealing. This is expected, since the sample is annealed far above the bulk Curie temperature of BaTiO$_3$ of about 120 °C where it exhibits a tetragonal to cubic phase transition and goes back to tetragonal phase when cooled where the domain alignment is reconfigured. The overall domain reorientation is illustrated in the histogram in Fig. 4c. Initially, the flake domains were aligned along multiple directions, but after annealing, the distribution narrows quite strongly around a single direction. This behaviour, however, does not appear to be systematic, as SHG measurements on a different flake (Supplementary Fig. S2) reveal a different trend.

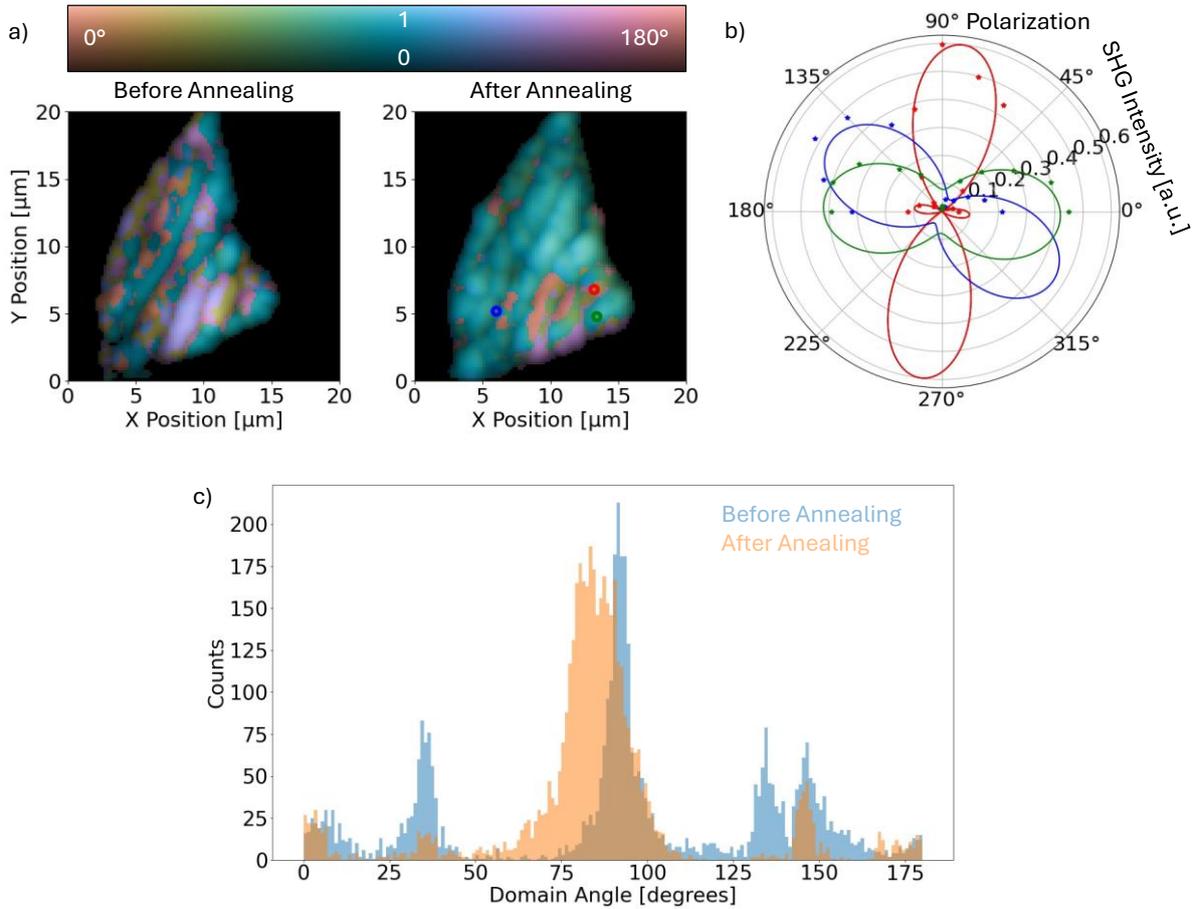

**Figure 4. SHG microscopy mapping of BaTiO$_3$ flake before and after annealing. a)** Polarization dependent SHG microscopy maps before (left) and after (right) annealing showing relative intensity encoded in brightness (0-1) and orientation of the domain polarization encoded in color (0°-180°). **b)** Exemplary polarization dependent SHG intensity in three selected points in a) (after annealing). Points show experimental data while solid lines show the fitting of the SHG intensity model presented in the Supplementary Information. **c)** Histogram showing the in-plane domain orientation before and after annealing, respectively.

Nonetheless, the reorientation patterns clearly indicate that the BaTiO$_3$ flakes processed via the CIS method already exhibit good structural and ferroelectric quality – even before the annealing step. At first glance, this appears to contradict the Raman spectroscopy results, which only show a significant recovery of the characteristic BaTiO$_3$ vibrational modes after annealing. However, this discrepancy can be understood by considering the nature of the Raman signal: it arises from lattice vibrations (phonons), which are particularly sensitive to point defects introduced during ion implantation. These defects can locally distort the lattice, strongly suppressing the Raman response. In contrast, the presence of long-range ferroelectric order – reflected in the domain reorientation behavior – can persist despite these defects. In other words, while the defects may disturb the precise phonon modes detectable by Raman spectroscopy, they do not necessarily destroy the macroscopic ferroelectric polarization in the material.

**Linear optical properties of BaTiO$_3$ flakes after defect curing**

To experimentally assess the linear optical response of the flakes, we performed reflectometry measurements across a spectral range of 500 to 950 nm on selected regions of a BaTiO$_3$ flake transferred onto SiO$_2$ on Si substrate (Fig. 5a). The ring-like features observed in the reflection are known as Newton's rings, which arise when a flat and a curved surface are in contact[30]. Their presence suggests that the exfoliated flakes are not entirely flat but are partially suspended above the $SiO_2$ surface. Measurements were taken at three distinct positions, indicated in Fig. 5a (see Methods for details). Reflection spectra are presented in Fig. 5b, indicating Fabry–Pérot oscillations in the flake within the measured spectral range.

The behavior of such multilayer optical systems can be quantitatively described using the Fresnel equations in combination with the transfer matrix method (TMM), which allows for modeling of interference effects in layered media. In the analysis, a three-layer system consisting of air, the $BaTiO_3$ thin film, and an underlying $SiO_2$ substrate was modeled (see Supplementary Information for details). Using literature values for the wavelength-dependent refractive index (dispersion) of bulk $BaTiO_3$[31], we calculated the reflectivity spectrum of a 1.2 µm-thick $BaTiO_3$ flake over the spectral range of 500 to 950 nm – parameters representative of our fabricated sample.

The TMM calculated reflectance spectra align closely with the measured spectra, indicating that the actual dispersion characteristics of the $BaTiO_3$ flake closely match those of the bulk material reported in literature. Deviations in the measured and calculated spectra can be attributed to a tiny and variable air gap between flake and underlying substrate as it can be concluded from the formation of Newtons rings.

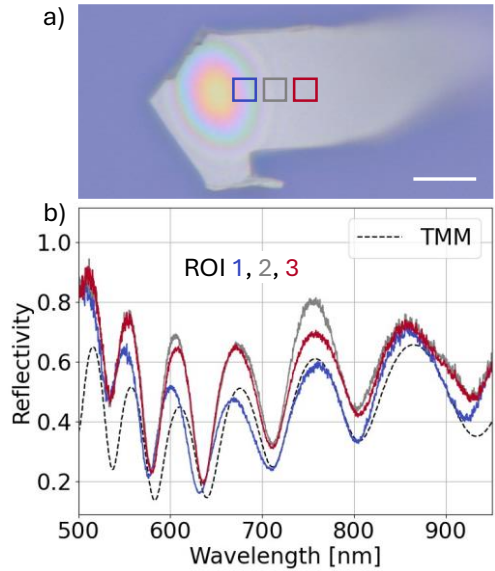

**Figure 5. Reflectometry of $BaTiO_3$ flake after annealing. a)** Bright field microscopy image of a $BaTiO_3$ flake, scale bar is 10 µm. **b)** Measured reflectivity in three different regions of interest (ROI) of the $BaTiO_3$ flake as indicated by the blue, grey and red rectangles in a). Red, grey and blue lines show the measured data while the black dashed line indicates the TMM modelling results for comparison.

**Conclusions and Outlook**

This work establishes Crystal Ion Slicing (CIS), combined with thermal annealing, as a qualified and scalable manufacturing route for high-quality $BaTiO_3$-on-insulator (BTOI) platforms. The study demonstrates that thermal annealing effectively mitigates the helium ion-induced damage inherent to the CIS process in $BaTiO_3$ thin films. Raman spectroscopy confirms the restoration of crystallinity, indicating that the structural degradation caused by ion implantation is largely reversible.

More critically, second-harmonic generation (SHG) microscopy reveals that annealing not only recovers the nonlinear optical properties but also enables reorientation of ferroelectric domains – directly influencing the $\chi^{(2)}$ tensor, which is essential for second-order nonlinear processes in photonics. The persistence of SHG signals in regions with low Raman intensity highlights an important decoupling between lattice coherence and long-range ferroelectric order, suggesting that BaTiO$_3$'s functional optical properties are surprisingly robust against ion-induced damage.

The observation that the linear optical dispersion of annealed CIS-sliced flakes matches that of bulk BaTiO$_3$ confirms the optical quality needed for photonic applications. This substantiates the suitability of CIS-processed BTOI films for integration into waveguides and nonlinear optical components.

Taken together, these results qualify CIS – augmented by appropriate annealing protocols – as a CMOS-compatible and scalable platform technology for BTOI manufacturing. This positions CIS-fabricated BTOI as a strong candidate for advanced photonic circuits, enabling high-speed modulation, efficient frequency conversion, and potentially integrated quantum light sources.

Looking forward, further refinement of the annealing process could unlock precise domain engineering within the BaTiO$_3$ layer, paving the way for programmable $\chi^{(2)}$ landscapes. This opens avenues for domain-based phase matching, reconfigurable photonic circuits, and on-chip entangled photon generation. Moreover, the compatibility of CIS-fabricated BTOI with established platforms like silicon and silicon nitride highlights its potential for hybrid photonic systems that integrate nonlinear, electro-optic, and quantum functionalities on a single chip.

In conclusion, this study not only validates the structural and optical recovery of CIS-processed BaTiO$_3$ films, but – more importantly – qualifies the CIS process as a viable manufacturing strategy for scalable BTOI platforms in next-generation photonic and quantum technologies.

## Methods

### Material Preparation and Ion Implantation

Two commercially available bulk BaTiO$_3$ single crystals, double-side polished, with (001) orientation were used as starting materials. The ion implantation was performed at room temperature using an air insulated 500 kV ion implanter with an indirectly heated cathode (IHC) ion source (Bernas type), a 40 kV extraction system and a 460 kV post acceleration unit. The extracted ions were separated by an analyzing magnet according to their q/A ratio and the 480 keV He$^+$ ion beam was scanned over the samples with a frequency of around 1 kHz in X- and Y-direction. Hereby the frequency of the X-direction is slightly different to that of the Y-direction to achieve a good homogeneity of the irradiation. To suppress ion channeling effects, the crystals were tilted by 7° off-axis during implantation. Two implantation fluences were employed: a low fluence of $5\times10^{16}$ ions/cm² and a high fluence of $2\times10^{17}$ ions/cm².

### X-Ray Diffraction

A Rigaku SmartLab X-ray diffractometer with a sealed Cu X-ray tube and a Ge (220) two-bounce monochromator was employed to measure 2θ/ω scans in the vicinity of the symmetric (002) and (200) reflections. The incident beam dimensions were set to 1 mm (incident slit) × 5 mm (length-limiting slit), with corresponding receiving slits and a HyPix detector in 0D mode.

### Thermal Processing and Flake Transfer

Following implantation, the BaTiO$_3$ crystals were annealed on a hot plate at 270 °C for 45 minutes to initiate exfoliation. The exfoliation and transfer process were carried out by placing the implanted BaTiO$_3$ surface in contact with a target substrate composed of 300 nm SiO$_2$ grown with plasma enhanced chemical vapor deposition on Si. This step facilitated the mechanical transfer of BaTiO$_3$ flakes from the bulk donor crystal to the receiver substrate.

After transfer, the BaTiO$_3$ flakes were subjected to a secondary annealing process in a conventional furnace. The samples were annealed in ambient air at 270 °C for 21 hours to promote crystallinity recovery and surface quality enhancement.

**Atomic force microscopy**

Atomic Force Microscopy (AFM) measurements were performed using a Dimension Edge system (Bruker), operating in tapping mode to minimize tip–sample interactions and preserve surface integrity during imaging. High-resolution topographic data were acquired with Tap300Al-G silicon cantilevers, specifically engineered for tapping mode, featuring nominal resonance frequencies around 300 kHz and force constants optimized for nanoscale surface profiling. Scans were conducted over a 20 × 20 μm² area to capture mesoscale morphological features, with images recorded at a resolution of 512 × 512 pixels to ensure high-fidelity surface reconstruction. A scan rate of 0.5 Hz was selected to achieve an optimal trade-off between acquisition speed, lateral resolution, and signal-to-noise ratio, enabling reliable analysis of fine surface features.

**Raman mapping**

Raman mapping was conducted using a Renishaw in Via Raman microscope (Renishaw plc, Gloucestershire, UK) equipped with a 473 nm excitation laser. The laser power at the sample surface was carefully adjusted to prevent sample heating and degradation. A 2400 lines/mm diffraction grating was employed to achieve high spectral resolution suitable for resolving fine vibrational features. Raman signals were collected through a 100× objective lens (numerical aperture ≈ 0.9), resulting in a lateral spatial resolution of approximately 1 μm.

The mapping was performed in point-by-point raster mode over a defined area, with a step size of 0.5 μm, yielding a total of 4920 spectra. Each spectrum was acquired with an integration time of 20 seconds and a single accumulation, providing high signal-to-noise data while preserving acquisition throughput. Silicon at 520.7 cm$^{-1}$ was used as a calibration standard to ensure wavenumber accuracy across the dataset. Spectral acquisition was conducted within the range of 100–1000 cm$^{-1}$, depending on the sample's vibrational features.

**Polarization-resolved second harmonic generation (SHG) microscopy**

The measurements were performed using a custom-built nonlinear microscopic setup. The fundamental excitation beam was provided by a femtosecond laser (Lightconversion Carbide, CB3-80W) delivering pulses of 211 fs duration at a repetition rate of 2000 kHz, with a tuneable wavelength range of 315–2600 nm through optical parametric amplification (Lightconversion Orpheus). The excitation wavelength was fixed at 900 nm for this measurement.

The laser beam was directed through a fixed linear polarizer and a half-wave plate (HWP) mounted on a rotation stage to control the incident polarization. The beam was then focused onto the sample using a long working distance objective (Mitutoyo Plan Apochromat, 0.42 NA, 50X). The sample was mounted on a XYZ piezoelectric stage (Physics Instruments) to ensure precise spatial alignment for the mapping.

The SHG signal generated in reflection was collected by the same objective and directed to a CMOS camera (Andor, Zyla 4.2), after reflecting from a long-pass dichroic mirror (cut-on wavelength at 650 nm) and passing through appropriate filters (two short-pass filters at 600 nm, two band-pass filters 335–610 nm and two short-pass filters at 850 nm) to suppress the fundamental laser light.

For polarization-resolved SHG mapping, the polarization state of the detected SHG signal was analyzed using a fixed linear polarizer perpendicular to the y-axis of the $BaTiO_3$ flake. The SHG intensity map was recorded while the analyzer was fixed and the input polarization was rotated via the HWP. The power of the excitation beam at the sample was maintained at 1.4 mW in front of the objective (800 μW at the flake/after the objective).

**Reflectivity measurements**

Reflectometry measurements were performed using a microscope (Zeiss) integrated with a spectrometer (Horiba, iHR320) with a cooled Si CCD detector. A Xenon short-arc light source with a broad spectrum (240–2400 nm) has been used to excite the sample via a 50x objective, and a small aperture has been applied to select only a small region of interest on the sample. The setting of the reflectometry setup was kept unchanged during the measurement. Then, a reference measurement with a protected silver mirror was carried out to normalize the reflectivity of the sample. The measurements were repeated several times to confirm the repeatability and intrinsic behavior of different positions on the sample but only three regions were shown in Fig. 5 for the ease of presentation.

# Supplementary Information:

# Structural and Optical Properties of Crystal Ion Sliced BaTiO$_3$ Thin Films

H. Esfandiar[1,2,*], F. Abtahi[2], T. G. Vrckovnik[1,2,3], G. Q. Ngo[1,2], R. Heller[4], U. Kentsch[4], F. Ganss[4], S. Facsko[4], U. Lucchesi[4], S. Winnerl[4], F. Eilenberger[1,2,3], D. Arslan[1], S. W. Schmitt[1,2,+]

[1] Fraunhofer Institute for Applied Optics and Precision Engineering IOF, Albert-Einstein-Str. 7, 07745 Jena, Germany

[2] Institute of Applied Physics, Abbe Center of Photonics, Friedrich Schiller University Jena, Albert-Einstein-Str. 15, 07745 Jena, Germany

[3] Max Planck School of Photonics, Albert-Einstein-Str. 15, 07745 Jena, Germany

[4] Institute of Ion Beam Physics and Materials Research, Helmholtz-Zentrum Dresden-Rossendorf, 01328 Dresden, Germany

[*]hossein.esfandiar@iof.fraunhofer.de

[+]sebastian.wolfgang.schmitt@iof.fraunhofer.de

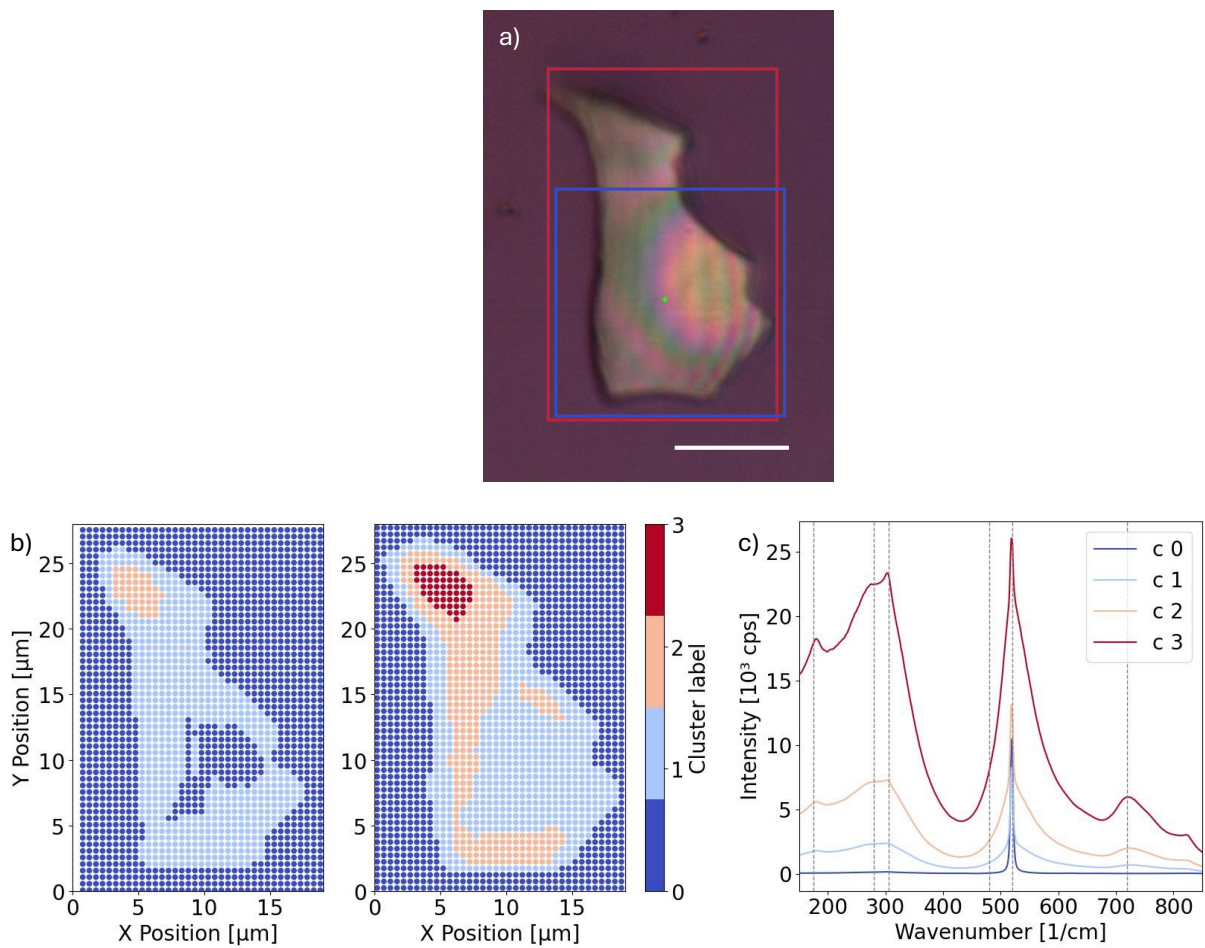

**Figure S1. Raman mapping of BaTiO$_3$ flake before and after annealing. a)** Bright field optical microscopy image of a BaTiO$_3$ flake exfoliated on a 300 nm SiO$_2$ thin film on Si. The red and blue frames show the area of the Raman and SHG microscopy scan in b), c) and Fig. S2, respectively. The scale bar is 10 μm. **b)** Clustered Raman maps before (left) and after (right) annealing of the BaTiO$_3$ flake shown in a). **c)** Cluster centroid spectra of the Raman maps shown in a). Dashed vertical lines show BaTiO$_3$ characteristic Raman peaks.

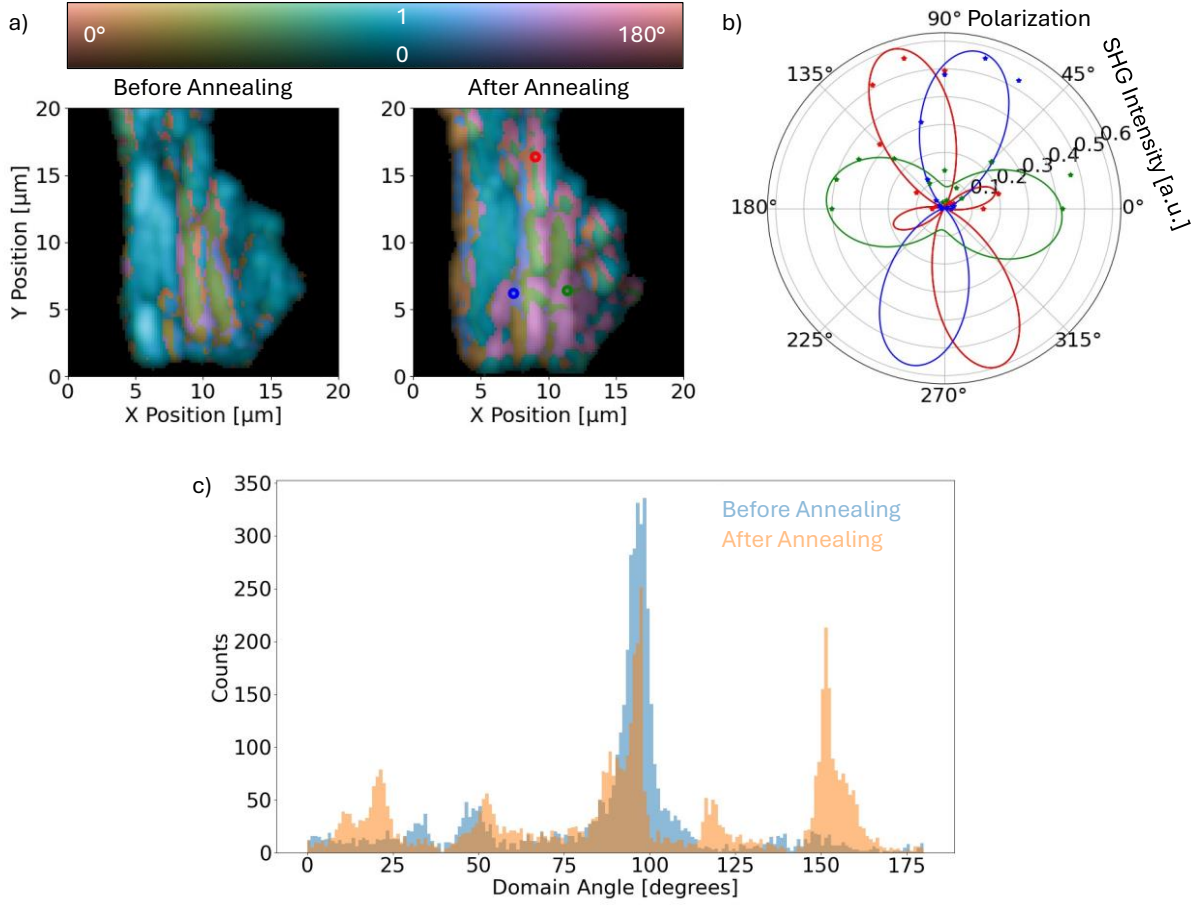

**Figure S2. SHG microscopy mapping of BaTiO$_3$ flake before and after annealing. a)** Polarization dependent SHG microscopy maps before (left) and after (right) annealing showing relative intensity encoded in brightness (0-1) and orientation of the domain polarization encoded in color (0°-180°). **b)** Exemplary polarization dependent SHG intensity in three selected points in a) (after annealing). Points show experimental data while solid lines show the fitting of the SHG intensity model presented in the Supplementary Information. **c)** Histogram showing the in-plane domain orientation before and after annealing, respectively.

**Fitting of the SHG microscopy data**

Assuming the crystal extraordinary axis mostly is in-plane, we define the angle between the lab x-axis and the extraordinary axis of the crystal as $\theta$, and the polarization of the incident electric field, E$_0$, relative to the lab x-axis as $\Phi$. Therefore,

$$\alpha = |\theta - \Phi| \quad (1)$$

is the relative angle between the crystal extraordinary axis and the polarization of the incident light at the fundamental frequency. In the crystal reference system, the incoming polarization can therefore be expressed as,

$$E_{x,c} = E_0 \cos(\alpha) \quad (2)$$
$$E_{y,c} = E_0 \sin(\alpha) \quad (3)$$

which assumes that the "x, c"-axis is parallel to the extraordinary axis of the crystal domain. This means the induced polarization from the crystal at the second harmonic polarization can be expressed as,

$$\begin{pmatrix} P_{x,c}(2\omega) \\ P_{y,c}(2\omega) \\ P_{z,c}(2\omega) \end{pmatrix} = \begin{pmatrix} d_{11} & d_{12} & d_{12} & 0 & 0 & 0 \\ 0 & 0 & 0 & 0 & 0 & d_{26} \\ 0 & 0 & 0 & 0 & d_{26} & 0 \end{pmatrix} \cdot \begin{pmatrix} E_{x,c}(\omega)E_{x,c}(\omega) \\ E_{y,c}(\omega)E_{y,c}(\omega) \\ E_{z,c}(\omega)E_{z,c}(\omega) \\ 2E_{y,c}(\omega)E_{z,c}(\omega) \\ 2E_{x,c}(\omega)E_{z,c}(\omega) \\ 2E_{x,c}(\omega)E_{y,c}(\omega) \end{pmatrix} \quad (4)$$

here for BaTiO$_3$ $d_{11}$ = 6.8 pm/V, $d_{12}$ = 15.7 pm/V, and $d_{26}$ = 17.0 pm/V[32]. Assuming the incoming light is purely polarized in the plane of the crystal (such that $E_{z,c}(\omega) = 0$), the above reduces to two simple equations,

$$P_{x,c}(2\omega) = d_{11} E_{x,c}^2(\omega) + d_{12} E_{y,c}^2(\omega) \quad (5)$$

$$P_{y,c}(2\omega) = 2d_{26} E_{x,c}(\omega)E_{y,c}(\omega) \quad (6)$$

Substituting in Equations (2) and (3) gives,

$$P_{x,c}(2\omega) = E_0^2[d_{11} \cos^2(\alpha) + d_{12} \sin^2(\alpha)] \quad (7)$$

$$P_{y,c}(2\omega) = 2d_{26} E_0^2 \sin(2\alpha) \quad (8)$$

The total induced polarization at the second harmonic frequency can then be found as,

$$P_c(2\omega) = \sqrt{P_{x,c}^2(2\omega) + P_{y,c}^2(2\omega)} \quad (9)$$

$$\beta = \arctan\left(\frac{P_{y,c}(2\omega)}{P_{x,c}(2\omega)}\right) \quad (10)$$

where $\beta$ represents the angle between the second harmonic polarization vector and the crystal extraordinary axis.

During measurement, the second harmonic light passes through an analyzer oriented parallel to the lab x-axis. Therefore, the polarization projected on the analyzer is,

$$P_{\text{an}} = |P_c(2\omega) \sin(\theta + \beta)| \quad (11)$$

The intensity measured by the detector is proportional to $P_{\text{an}}^2$. Therefore, given an input electric field, $E_0$, polarization angle, $\phi$, and crystal domain angle, $\theta$, the expected measurement after the analyzer can be found by following the above steps. Conversely, knowing the output intensity of the second harmonic light given the input polarization angle allows us to find the orientation of the crystal domains. To do this, we use a normalized cross-correlation (NCC) method. Here we first calculate the output intensities, $I_{\text{out}} = P_{\text{an}}^2$ for each value of $\theta$ from 0° to 180° in 1° steps, for the 13 different input polarization angles ($\phi$) used for the measurements, assuming $E_0 = 1$. Then the NCC is calculated as,

$$\text{NCC}(\theta) = \langle \frac{I_{\text{meas}}(\Phi)}{|I_{\text{meas}}(\Phi)|}, \frac{I_{\text{out}}(\Phi, \theta)}{|I_{\text{out}}(\Phi, \theta)|} \rangle \quad (12)$$

where $I_{\text{meas}}$ is the experimentally measured intensities as a function of the input and $\langle .,. \rangle$ represents the inner product. This gives a measure between 0 and 1 indicating how similar the experimental spectrum is to each of the theoretical spectra with the crystal in different orientations. The normalization makes it such that the values of $E_0$ are irrelevant for this comparison. We then take the true value of $\theta$ to be that which maximizes the NCC,

$$\theta = \arg\max(\text{NCC}(\theta)) \quad (13)$$

To get an approximation of the SHG intensity, we then see by what value the theoretical spectrum (calculated with $E_0 = 1$), $I_{\text{out}}$ needs to be scaled to best match the measured spectrum. This is found as,

$$I_{\text{SHG}} = \frac{\langle I_{\text{meas}}, I_{\text{out}} \rangle}{\langle I_{\text{out}}, I_{\text{out}} \rangle} \quad (14)$$

This derivation allows us to map out the full flake and record both the expected crystal domain orientation and SHG intensity, as seen in Fig. 3d)[33,34].

NCC maps for the four SHG maps presented in Figs. 4 and S2 are shown in Fig. S3. Note that a lower NCC seems to occur predominantly at domain boundaries, as here domain orientation is not unambiguously defined within the measurement resolution. Accordingly, the NCC can be associated with a map of the domain boundaries.

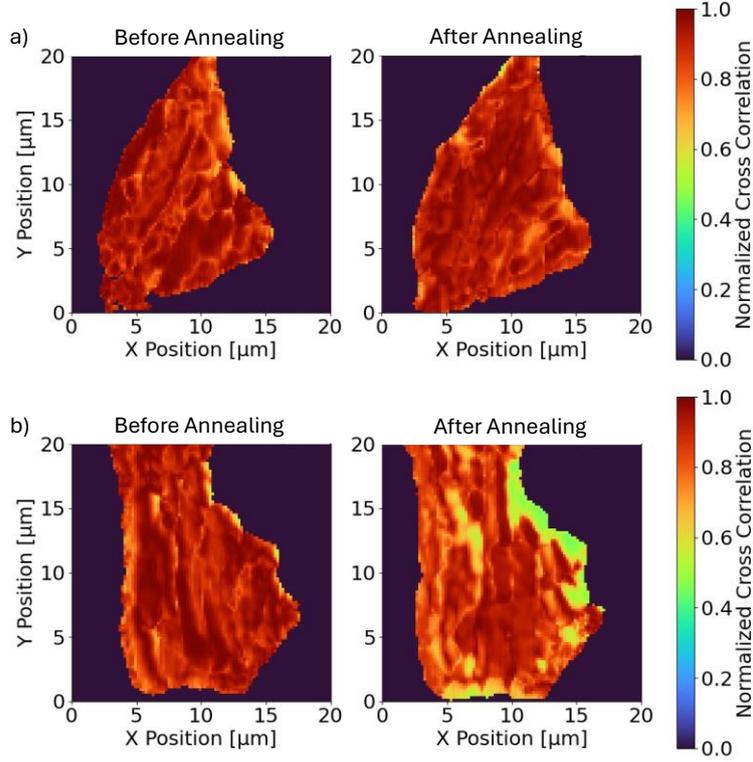

**Figure S3. Fitting quality for polarisation-resolved SHG mappings a)** NCC for SHG mapping of $BaTiO_3$ flake shown in Fig. 4 before and after annealing. **b)** NCC for SHG mapping of $BaTiO_3$ flake shown in Fig. S2 before and after annealing.

**Reflectivity calculation**

The reflectivity was calculated as:

$$R = \left| \frac{(k_1 M_{22} - k_3 M_{11}) - i(M_{21} + k_1 k_3 M_{12})}{(k_1 M_{22} + k_3 M_{11}) + i(M_{21} - k_1 k_3 M_{12})} \right|^2 \tag{15}$$

$k_1$, $k_2$, $k_3$ are the wavevectors in the respective media (air, $BaTiO_3$, and $SiO_2$ substrate, respectively).

The refractive index of $BaTiO_3$ was taken from literature using a dispersion formula[31]. The refractive index of $SiO_2$ was taken from the Sellmeier equation[35].

Transfer Matrix Elements ($M_{ij}$) describe how light propagates through $BaTiO_3$ thin film and the response at the interfaces. The elements of the transfer matrix are given by:

$$M_{11} = \cos(k_2 d) \tag{16}$$

$$M_{12} = \frac{1}{k_2} \sin(k_2 d) \tag{17}$$

$$M_{21} = -k_2 \sin(k_2 d) \tag{18}$$

$$M_{22} = \cos(k_2 d) \tag{19}$$

with d being the thickness of the $BaTiO_3$ flake.